\documentclass[11pt,conference]{IEEEtran}
\usepackage[T1]{fontenc}
\usepackage{graphicx}
\graphicspath{{Figures/EPS/}}
\usepackage{setspace}
\usepackage{amssymb,textcomp, amscd,amsmath,amsfonts,mathrsfs}
\usepackage{subfigure}
\usepackage{verbatim}
\usepackage{float}
\usepackage{balance}
\usepackage{hyperref}

\IEEEoverridecommandlockouts
\begin{document}

\title{Recent Advances in 3GPP Rel-12 Standardization related to D2D and Public Safety Communications}

\author{\IEEEauthorblockN{Dorin Panaitopol*\thanks{*This work has been initially submitted to WCNC 2014 Workshop on Device-to-Device and Public Safety Communications (WDPC). Dorin Panaitopol is now working with Thales Communications \& Security (TCS), WaveForm Design (WFD) Laboratory, 4 Avenue des Louvresses, 92622 Gennevilliers Cedex, France. He can be contacted at Dorin.Panaitopol@thalesgroup.com}, Christian Mouton, Benoit Lecroart, Yannick Lair and Philippe Delahaye \\(previously with NEC Technologies (UK) Ltd., FMDC Lab., ComTech Department)} 
\IEEEauthorblockA{
Email: Dorin.Panaitopol@thalesgroup.com\\
}
}

\maketitle
\begin{abstract}

The goal of this paper is to present advances on recent 3GPP standardization activities related to Device-to-Device (D2D) and public safety. The paper provides a clear 3GPP state of the art, including when the 3GPP work on D2D and public safety communications started. Finally, it presents important conclusions with respect to further 3GPP work on this topic.

\end{abstract}


\section{Introduction}

In the past few years D2D communication emerged as an important research topic and many publications provided D2D communication advantages over LTE such as higher data rates and better resource reallocation \cite{Nokia, Ericsson}. Nowadays D2D communication for LTE technology is sufficiently mature that it can be standardized \cite{Waterloo}. Moreover, recently a strong move has been made by the public safety ecosystem, starting in the US, towards LTE and 3GPP for the development of next generation of public safety networks, as shown in \cite{Paulson}. 

\begin{figure}[htbp]
\centering
\includegraphics[scale=0.3]{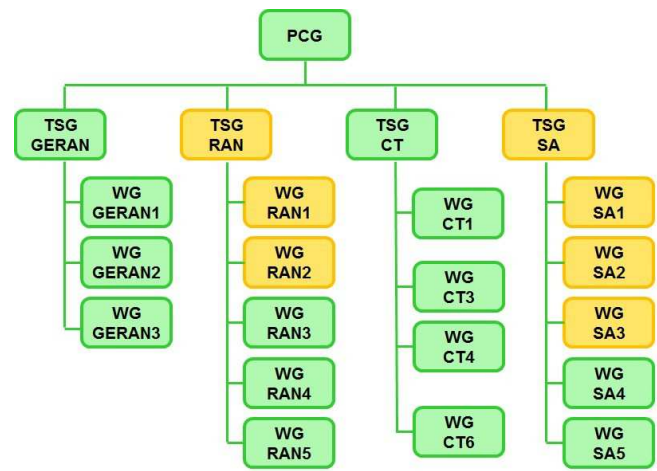} 
\caption{Structure of 3GPP and D2D active groups as in June 2014}
\label{3GPP}
\end{figure}

In order to better understand the evolution of 3GPP standardization activities for Device-to-Device (D2D) communications and how this subject relates to public safety in the latest releases, the next paragraphs describe the structure and the organization of 3GPP. As depicted in Fig.~\ref{3GPP}, the highest decision making group in 3GPP is the Project Coordination Group (PCG) being for example responsible for final adoption of 3GPP Work Items (WIs), ratifying election results and resources committed to 3GPP. Below the PCG there are four Technical Specification Groups (TSGs): 
\begin{itemize}
	\item GERAN (GSM EDGE Radio Access Networks), responsible for the radio access specification of GSM/EDGE, 
	\item RAN (Radio Access Networks), responsible for the definition of the functions, requirements and interfaces of UTRA/E-UTRA network, 
	\item CT (Core Network \& Terminals) responsible for specifying terminal interfaces (logical and physical), terminal capabilities (such as execution environments) and the core network part of 3GPP systems,
	\item SA (Service \& Systems Aspects), responsible for the overall architecture and service capabilities of systems based on 3GPP specifications.  
\end{itemize}

As presented in Fig.~\ref{3GPP}, each of the four TSGs is structured in different working groups, each 3GPP Working Group (WG) having a particular area of expertise, as will be explained later on. Moreover, TSG SA has the responsibility for cross TSG coordination. Therefore, as shown in the next paragraphs, each new 3GPP WI is first initiated by TSG SA. Obviously, this was also the case for D2D communications. 

One of the responsibilities of 3GPP is to produce Technical Reports (TRs) and Technical Specifications (TSs) on evolution of existing radio technologies or new radio technologies defined by 3GPP, such as HSPA/HSPA+, LTE and LTE-Advanced (LTE-A). 3GPP specifications are grouped into "Releases", and a mobile system can be implemented based on the set of all specifications which comprise a given Release. Whenever a new feature is required by the market, it is proposed to be specified in a given release. Currently 3GPP is working on LTE-A Rel-12, intended to be functionally frozen in June 2014. Two of the major features to be developed in Rel-12 are D2D communication and discovery, as a support for so-called Proximity Services (or ProSe). These services are considered for use in commercial scenarios (e.g. for providing higher data rate or reuse of radio resources locally, based on proximity of communicating users) and in scenarios related to public safety and critical communications.

Currently, from Fig.~\ref{3GPP}, only the highlighted groups are involved. WG SA1 (working group responsible for service requirements) started to work on a feasibility study on D2D ProSe at the end of 2011, with the outcomes compiled in TR $22.803$ \cite{22803}. Normative work was then conducted by WG SA1, mainly in TS $22.278$ \cite{22278}. Based on these normative requirements, several working groups are also conducting a feasibility study, WG SA2 for the architecture, WG SA3 for the security aspects of proximity-based services. For example, WG SA2 possible solutions have been compiled in TR $23.703$ \cite{23703}. In parallel, WG RAN1 (working group responsible for the specification of the physical layer of the radio interface), and to some extent, WG RAN2 (in charge of the radio interface architecture and protocols), are evaluating the feasibility of providing D2D ProSe via LTE in TR $36.843$ \cite{36843}.

\begin{figure}[htbp]
\centering
\includegraphics[scale=0.155]{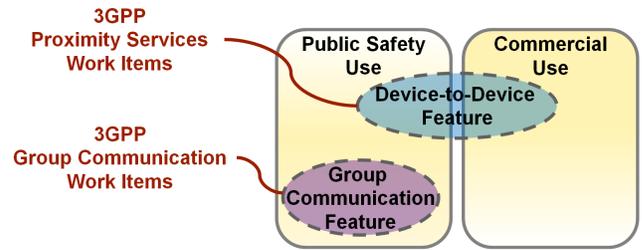} 
\caption{3GPP Proximity Services and Group Communication Work Items}
\label{3GPP_WIs}
\end{figure}

In parallel with ProSe WI, 3GPP has started standardization work for Group Communication System Enablers for LTE (GCSE\_LTE) designed for critical communications purposes, leading to the requirement for groupcast (or 1-to-many) communications to be supported by ProSe. It has therefore been decided that the purpose of GCSE\_LTE WI is not for the time being for commercial use but only for public safety use as explained in Fig.~\ref{3GPP_WIs}.

For the reasons mentioned above, Table~\ref{WID_bis} summarizes the current technical reports and specifications not only for ProSe but also for GCSE\_LTE scenarios because both represent public safety scenarios. As presented in Table~\ref{WID_bis}, SA1 TS $22.468$ \cite{22468} and SA2 TR $23.768$ \cite{23768} are the main documents of interest for GCSE\_LTE.

\begin{table*}[!t]
\caption{\label{WID_bis} 3GPP Current Specifications and Work Items on D2D}
\begin{centering}
\begin{tabular}{|l|l|l|l|l|}
\hline 
 Responsible & Related & TS or TR & Number &Status (as at October 2013) \tabularnewline
Work Group & Work Item  &  & &  \tabularnewline
\hline
\hline
 SA1 & ProSe & TS & $22.278$ \cite{22278} & Rel-12 requirements are frozen \tabularnewline
\hline
 SA1 & GCSE\_LTE & TS & $22.468$ \cite{22468} & Rel-12 requirements are frozen \tabularnewline
\hline
 SA2 & ProSe & TR & $23.703$ \cite{23703} & Document is being progressed in 3GPP \tabularnewline
\hline
 SA2 & GCSE\_LTE & TR & $23.768$ \cite{23768} & Document is being progressed in 3GPP \tabularnewline
\hline
 RAN1 & ProSe & TR & $36.843$ \cite{36843} & Document is being progressed in 3GPP \tabularnewline
\hline
\end{tabular}
\par\end{centering}
\end{table*}

The rest of this paper is organized as follows. Section \ref{Sec_UseCases} presents the key use cases for 3GPP D2D related technologies and summarizes the D2D required features in Rel-12. Some generalities about ProSe architecture are then described in Section \ref{Sec_CoreOptions}. In relationship with Section \ref{Sec_CoreOptions}, an important attention is dedicated to discovery feature in Section \ref{D2D_Discovery} and to communication feature in Section \ref{D2D_Communication}. Finally, 3GPP ProSe roadmap for Rel-12 and important conclusions are provided in Section \ref{Sec_3GPPOrg} and Section \ref{Sec_Conclusion} respectively.

\section{Use Cases for D2D Technologies in 3GPP and Required Features}
\label{Sec_UseCases}

The standardization work on D2D technologies in 3GPP is focused on a set of use cases, which were identified to fit the needs of both public safety and commercial mobile networks. Use cases can be defined by various services offered in various situations. The core features supported by D2D technologies in 3GPP are:
\begin{enumerate}
	\item direct discovery;
	\item direct 1:1 communication;
	\item direct 1:many communication.
\end{enumerate}

Here "direct" means making use of the direct radio interface between the devices instead of going through the network infrastructure, this use being under the control of the network operator, either through on-line (i.e., making use of cellular links of the users) or off-line (i.e., pre-configuration of User Equipments (UEs)).

The direct discovery feature is designed to support a new discovery service, to be offered to users to "discover" other users in the vicinity. Note that some Enhanced Packet Core (EPC) based solution (i.e., without use of the D2D interface) is also considered to offer this service, with the limitation that it would only work on-line. The discovery can therefore be an EPC-level ProSe discovery (if the discovery is performed by the EPC) or a ProSe direct E-UTRA discovery (if discovery is performed at radio level).


The direct 1:1 communication feature is designed to support the usual data communication service between 2 users. The direct 1:many communication feature is designed in order to support a new groupcast service called Group Communication (inspired by GCSE\_LTE WI and which is not covering the direct communications aspects 1:1 or 1:many). From the point of view of direct communication, the UEs may communicate directly with each other via LTE technology, and the communication may happen with or without network assistance (e.g., signaling and control) for both 1:1 and 1:many situations. Direct D2D communication via WLAN connection is also possible with the support of the EPC.



On top of direct communication, 3GPP defines a UE relaying feature which is used for public safety scenarios. This feature is applicable in situations such as UE-to-Network relaying (when a UE is relayed to the network by a UE with relaying capabilities called UE-Relay) or UE-to-UE relaying (when a UE is relayed to another UE with the help of a UE-Relay). To some extent this can be already included to the direct communication features. However, this can be seen as a different feature at least from the point of view of the control: a UE relaying another UE could also be able to perform e.g. Radio Resource Control (RRC) functions that normally a legacy UE cannot support by itself.


With respect to all core features, and more precisely features that a D2D UE and a D2D-supporting network should have, Table~\ref{Req_Features} summarizes the current Rel-12 3GPP priority per WI. 


\begin{table}[htbp]
\caption{\label{Req_Features} Required Features for ProSe in Rel-12}
\begin{centering}
\begin{tabular}{|l|l|l|}
\hline 
D2D & Public Safety & Commercial  \tabularnewline
Feature & Use Case    & Use Case   \tabularnewline
\hline
\hline
Discovery & Required & Required  \tabularnewline
          & (on-\&off-line) & (on-line) \tabularnewline
\hline
1:1 & Required & Required  \tabularnewline
Communication & (on-\&off-line) & (on-line)   \tabularnewline
\hline
1:Many & Required & Not Required  \tabularnewline
Communication & (on-\&off-line) &   \tabularnewline
\hline
with UE & Required & Not Required  \tabularnewline
Relaying& (on-\&off-line) & \tabularnewline
\hline
\end{tabular}
\par\end{centering}
\end{table}

For each feature, multiple use cases can be described for a pair of users (then they can be extrapolated to any number of users) depending on:
\begin{enumerate}
	\item whether none, one or both users are served by a D2D-supporting network;
	\item when both are served by adequate network, whether it is by the same network or two different networks;
	\item when both are served by same network, whether they are served by the same cell or different cells;
	\item whether both users are subscribers of the same operator;
	\item whether none, one or both users are in a roaming situation.
\end{enumerate}
 
Moreover, these core features described above are subject to different level of performance with regard to:
\begin{enumerate}
	\item management of Quality of Service (QoS);
	\item management of security;
	\item management of service continuity.
\end{enumerate}

QoS, security and service continuity are therefore key performance indicators that have to be taken into account by 3GPP work on D2D communications.

\section{Core Options for System Architecture}
\label{Sec_CoreOptions}

At the beginning of 2013 the SA2 group started the system architecture definition suitable to fulfill the requirements identified for ProSe and GCSE. As usual for defining complex new features, a two-step approach was performed. In a first step, a TR document proposing different solutions would be defined. In a second step, the best solution would be chosen, and would lead to the definition of TS documents that will be the normative specification of the features. Since the beginning of 2013, TR $23.703$ \cite{23703} and TR $23.768$ \cite{23768} have progressed and currently contain the following information: 
\begin{itemize}
	\item a list of key issues to be solved;
	\item the definition of an overall system architecture;
	\item the identification of technical solutions expected to meet the system requirements.
\end{itemize}

\begin{figure}[htbp]
\centering
\includegraphics[scale=0.2]{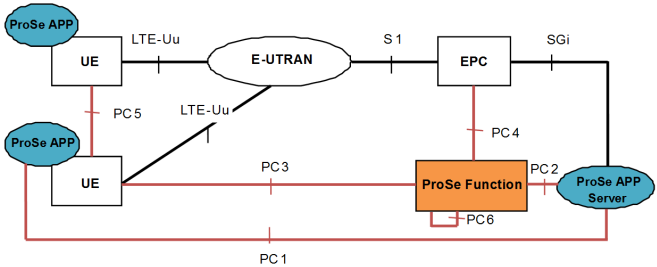} 
\caption{ProSe Overall Architecture}
\label{architecture}
\end{figure}

An overall architecture for the non-roaming cases, and a variant for the roaming cases were defined. Those architectures identified the main components and reference points of the system. For example, ProSe overall architecture \cite{23703} for the non-roaming case is presented in Fig.~\ref{architecture}.

As described by Fig.~\ref{architecture}, six reference points have been introduced, as well as a ProSe Function located in the EPS, and a ProSe APP Server that is outside of the EPS. The new component "ProSe Function" is created to support the Evolved Packet System (EPS) features of ProSe. The ProSe APP Server, which is located outside the 3GPP network, represents the application server of a ProSe provider. This architecture has to support a wide range of solutions and is thus very generic and flexible.

The overall architecture has also to support all the features of ProSe, such as the discovery and the communication between UEs. This later one includes the communication required allowing a UE out-of-network coverage to access to the network infrastructure through a UE-Relay.

The architecture allows a wide split of functionalities and a different type of interfaces. The ProSe Function and the ProSe APP Server split can be very different. Some solutions can be based on the use of IMS in the ProSe APP Server, other solutions allow to use ProSe control plane messages over the PC3 interface between the UE and the ProSe Function by using IP user plane messages, while other solutions could use Non-Access Stratum (NAS) messages.

Another architecture has been defined in $23.768$ \cite{23768} for GCSE\_LTE. This architecture has a new component GCSE Application Server (GCSE AS) and a Multipoint Service (MuSe) component most probably based on already existent Multimedia Broadcast/Multicast Service (MBMS)-type solution. In the case of GCSE\_LTE, five new reference points have been proposed: GC1 between GCSE AS and GCSE application on UE side, GC2 between the GCSE AS and MuSe, GC3 between MuSe and the base station eNB, GC4 between MuSe and the Mobility Management Entity MME, and finally GC5 between MuSe and the Packet-GateWay (P-GW).

In ProSe TR 23.703 \cite{23703}, the solutions proposed to meet the ProSe requirements can be divided in four categories:
\begin{itemize}
	\item solutions related to the definition of ProSe identities (see Section \ref{D2D_Discovery} and more precisely Subsection \ref{Identifiers}); 
	\item solutions related to direct discovery (see Section \ref{D2D_Discovery} and more precisely Subsection \ref{Disc_Information_in_RRC});
	\item solutions related to direct communication (see Section \ref{D2D_Communication}); 
	\item solutions related to UE-Relay.
\end{itemize}

The next sections \ref{D2D_Discovery} and \ref{D2D_Communication} provide a review of the latest 3GPP RAN2 \cite{RAN283bis} and SA2 \cite{SA299} discussions and the material provided in these sections tries to summarize the most important contributions.

\section{3GPP D2D Discovery Solutions}
\label{D2D_Discovery}
 
As previously mentioned, the first subsection from Section \ref{D2D_Discovery} (i.e. Subsection  \ref{Identifiers}) deals with the UE and Group Identities for both discovery and communication on PC5, while the second subsection (i.e. Subsection \ref{Disc_Information_in_RRC}) shows how the discovery information is being transmitted and received. 

\subsection{User Equipment and Group Identities}
\label{Identifiers}

\begin{figure}[htbp]
\centering
\includegraphics[scale=0.135]{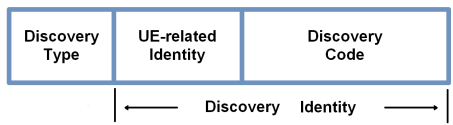}
\caption{Example of Discovery Message}
\label{identities}
\end{figure}

In order to perform the direct discovery, a UE has to broadcast a discovery information using direct communication. The direct 1:1 communication could be done in a Question/Answer-based manner (e.g. Question: Is the service or the UE "A" around? followed by the Answer: Yes, "A" is available at UE-John.Doe) or in an Announcement-based manner which is regularly repeated during the broadcast period (e.g. Announcement: Service "A" is available at UE-John.Doe). The information communicated during the D2D communication therefore contains a discovery identifier relating D2D communication to the UEs and to the service (i.e. "Discovery Type" from Fig.~\ref{identities}). After the discovery identities are defined and produced by e.g. an application server (or by a UE), they have to be communicated to the UEs that need to use them either for broadcast purposes, either for detection purposes. In the particular cases of public safety where UEs are out-of-network coverage, the discovery identities have to be independently determined by each concerned UE (i.e. "UE-related Identity" from Fig.~\ref{identities}).

More than that, the discovery identities need to be unique in all circumstances. In particular, it has to be noted that the discovery service can be performed in an inter-operator scenario or when UEs are in a roaming situation, and therefore the UE-related Identity may not assure uniqueness. In order to keep the uniqueness of the identities, some proposed solutions use an application-level URL-type identifier (e.g., John.Doe@identities.myapplication.com) that can be converted to a unique discovery code (i.e. "Discovery Code" from Fig.~\ref{identities}) to be broadcasted between the UEs. Some other solutions add Layer 2 identities of discoveree/discovered UEs, i.e. a solution proposed to create a generic Layer 2 source and destination addresses in each D2D discovery messages. 

For exemplification, Fig.~\ref{identities} presents a possible example of discovery message composed from: the type of discovery message, the UE-related identity and the discovery code.

Moreover, in the particular case of public safety UEs, the discovery is used to find UEs related to a particular communication group. A specific identity related to the group could be therefore added or included as part of the application identifier (e.g. John.Doe:Group-12@brigade13-fireman.co.uk). Also, for public safety UEs, the discovery services could be available when the UEs are out-of-network coverage requiring a pre-configuration of identifiers or different methods for local generation of identifiers. With respect to all the above mentioned, the security of the identifiers broadcasted over the air will be defined by the 3GPP WG SA3 in order to keep the required level of confidentiality and security.

\subsection{Discovery Information Transmission and Reception}
\label{Disc_Information_in_RRC}

The goal of Subsection \ref{Disc_Information_in_RRC} is to describe how the discovery information is transmitted (see Subsection \ref{TxDiscInfo}) and how discovery information is received and used (see Subsection \ref{RxDiscInfo}) for different RRC states.

\subsubsection{Transmission of Discovery Information}
\label{TxDiscInfo}

In current LTE network system, legacy UE transmission control by eNB during RRC connected mode ensures that eNB dynamically restricts the intra-cell and inter-cell interference. However, the eNB does not need to control RRC idle mode legacy UEs since they do not cause interference. As explained below, the needs are different when considering direct discovery because 1) in RRC idle mode the D2D UEs may need to transmit without directly entering in RRC connected mode and 2) too many transitions from the RRC idle mode to RRC connected mode are expensive from the signaling point of view.

For example, users running an application on their phone may want to find other nearby users of the same application at any time. Requesting the UE to enter RRC connected mode to transmit a discovery message and then to enter again RRC idle mode once the message has been transmitted would result in a high signaling overhead. It is therefore desirable that a UE transmitting D2D discovery messages be allowed to transmit while in RRC idle mode. Moreover, the transition of a UE from RRC idle to connected mode may result in random access procedure, RRC connection establishment, initial security activation, default EPS and radio bearer establishments. Thus, any RRC idle to connected mode transition caused by discovery will overload the random access procedure, will increase the control plane overhead and will impact the UE power consumption.  

\subsubsection{Reception of Discovery Information}
\label{RxDiscInfo}

It seems obvious that allowing devices to receive discovery messages while in RRC idle mode would provide several advantages compared to requesting the devices to enter RRC connected mode. 

For example, a user can activate an application which is notifying of the proximity of restaurants of interest. The relevant application would conceivably be continuously running on the device, meaning that if discovery was only permitted while in RRC connected mode, the device would be unable to leave this state unless the application was deactivated. However, requesting a UE to remain in RRC connected mode to receive discovery messages may be unacceptable for several reasons. Firstly, the power consumption of a UE remaining in RRC connected mode may be considerably high. Secondly, compared with a UE in RRC idle mode, a UE remaining in RRC connected mode may increase the amount of signaling throughout the network as a result of handover procedures. It is therefore much more practical for the device to discover a restaurant while in RRC idle mode and only enter RRC connected mode if the user requested more information on the restaurant, such as opening times and menus.

\section{3GPP Radio Resource Allocation Schemes for D2D Communication}
\label{D2D_Communication}

The main focus of Section \ref{D2D_Communication} is on direct 1:many communication for 3GPP Rel-12 for in-network coverage and out-of-network coverage situations. In order to ensure efficient usage of resources and to allow an efficient UE power consumption in both situations, two main types of resource allocation schemes have been currently proposed: 
\begin{itemize}
	\item Centralized approach (see Subsection \ref{Centralized}): The network controls the resource allocation. This approach can either apply to an eNB in-network coverage or to a cluster head in out-of-network coverage case. The centralized approach consists in a dynamic or semi-persistent scheduling for which the UE is scheduled with resources for every D2D transmission.
	\item Distributed approach (see Subsection \ref{Distributed}): The UE autonomously allocates resources by itself from a pool of semi-statically configured resource blocks. As the resource allocation is distributed, and therefore same resource blocks can be allocated at the same time by different UEs, there might be a contention situation that has to be properly solved.
\end{itemize}

\subsection{Centralized Approach}
\label{Centralized}

For the centralized approach, the D2D transmitting UE sends a scheduling request to the network controller or centralized node. Upon receiving the scheduling request from the transmitting UE, the centralized node allocates a resource which is not being used by other member UEs (i.e. UEs involved in groupcast communication) or by all UEs (i.e. UEs in proximity) within the cell or within the coverage of the centralized node. The centralized node assigns resources based on the amount of data to be sent and also based on the radio conditions. If the amount of data requested by the transmitting UE cannot be fulfilled by a single schedule, multiple schedules are required. Depending on the application or the type of service, it may also be possible to provide semi-persistent resources. The centralized node also informs the D2D UEs with the resources used for transmission by other member UEs involved in a groupcast communication or all other UEs which are in proximity.

\subsection{Distributed Approach}
\label{Distributed}

For the distributed approach, a pool of resources is semi-statically allocated by the network (i.e. for the in-network coverage case) or is pre-configured when the UE is registered with the network (i.e. for the out-of-network coverage case). The UEs use this pool of resources to transmit/listen direct discovery beacons, but the pool of resources can also be used for direct 1:1 and 1:many communication. As it is a contention-based access, some form of collision handling mechanism (e.g. Carrier Sense Multiple Access Collision Detection or Collision Avoidance, CSMA/CD or CSMA/CA) is needed to detect/avoid/resolve any collision. In CSMA, the transmitting UE listens to the pool of resources to determine which of resources are being used or not for transmission by other UEs. Once it determines that a resource is not occupied, the UE may transmit using the same resource. The random selection of the unused resources may decrease the number of collisions, but collisions may still occur if the pool of resources is limited. 

\subsection{Centralized Approach vs. Distributed Approach}

In Subsection \ref{Centralized} and Subsection \ref{Distributed} two main approaches were presented: centralized approach (i.e. full scheduling) and distributed approach (i.e. CSMA-like resource allocation). Table~\ref{Approaches} further provides a comparison of the two main approaches, with their advantages and disadvantages.

\begin{table}[ht]
\caption{\label{Approaches} Comparisons of the Resource Allocation Approaches}
\begin{centering}
\begin{tabular}{|l|l|l|}
\hline 
 & Centralized & Distributed \tabularnewline
 & Approach & Approach \tabularnewline
\hline
\hline
Resource & High & Low (depending on the \tabularnewline
Efficiency &  & type of CSMA solution) \tabularnewline
\hline
Signaling & High & Low \tabularnewline
Overhead &  &  \tabularnewline
\hline
Pre-configuration & No & Yes (e.g., determination \tabularnewline
of Resources &  & of resource pool) \tabularnewline
\hline
Interference & Low & Medium \tabularnewline
\hline
\end{tabular}
\par\end{centering}
\end{table}

\begin{figure*}[ht!]
\centering
\includegraphics[scale=0.29]{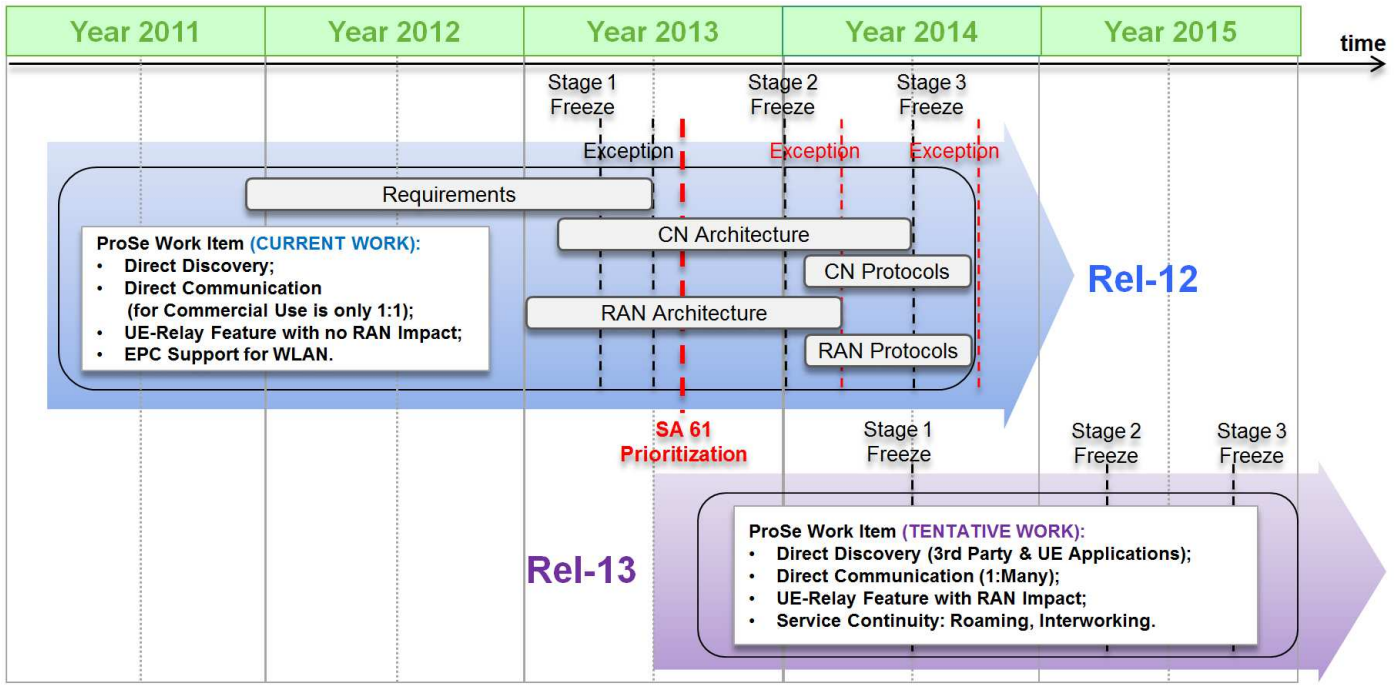} 
\caption{3GPP Prose Tentative Roadmap for Rel-12 and Rel-13}
\label{Roadmap}
\end{figure*}

Based on Table~\ref{Approaches}, it can be easily observed that CSMA-like mechanism has relatively lower complexity and signaling overhead. However, the drawback of the distributed approach is its weakness in terms of interference control and collision resolution. The collision resolution is typically implemented using probabilistic approaches, e.g. using medium/carrier sensing and a random back-off timer to perform retransmission if medium is occupied (e.g. CSMA/CA). These mechanisms typically increase power consumption of the UE when active and may lead to degradation of system performance. On the other hand, centralized approach has the advantage that it can efficiently manage radio resources in a cell and it can avoid UE interference with other UEs or with the network. For D2D communications, the centralized approach may also need to consider the D2D channel condition and the D2D buffer status reported by the UEs. If for example the number of transmitting D2D UEs is high, this will probably increase the signaling overhead significantly and therefore new measurement, configuration and reporting methods have to be defined by 3GPP.

\section{3GPP ProSe Roadmap and Prioritisation}
\label{Sec_3GPPOrg}

Among other important activities, 3GPP TSG RAN and 3GPP TSG SA are also involved in D2D proximity services from overall work plan management. In September 2013, during SA 61 meeting, 3GPP TSG SA reviewed the Rel-12 progress in different working groups. Due to the important activities in the radio access area, not only due to proximity services, 3GPP TSG RAN proposed to reduce the scope of proximity services in Rel-12. Similarly, experimenting an overload situation, 3GPP WG SA2 proposed to limit Rel-12 to a list of essential features. Based on these information, 3GPP TSG SA decided to limit the proximity features for Rel-12 to discovery and public safety group communication.

3GPP ProSe tentative roadmap for Rel-12 and Rel-13 can be found in Fig.~\ref{Roadmap}. The UE-Relay feature was maintained in Rel-12, but is considered as a lower priority item and remains at risk for the final release. Regarding the overall schedule for Rel-12, it is highly possible to have a planning exception and complete three months later than the official timescale currently set to June 2014. The service aspects and network requirements (covered by stage 1) are now complete, but the completion of the architecture (stage 2) would then be finalized by March 2014 and the completion of the protocols (stage 3) would be finalized by September 2014 for both Core Network (CN) and RAN-related aspects. With respect to Rel-13, the main work will probably concern direct discovery feature (including interaction with 3rd party applications and UE terminal applications), direct 1:many communication feature for commercial use, UE-Relay feature with RAN impact (not covered by Rel-12) and service continuity aspects.

\section{Conclusion}
\label{Sec_Conclusion}

This paper shows the latest 3GPP contributions on D2D communications and public safety. With respect to all the D2D features presented above (i.e., direct discovery, direct 1:1 and 1:many communications) there are some important research topics that have to be considered in order to improve the user experience such as energy consumption, QoS, security, resource allocation, interference management (i.e., including interference avoidance) and service continuity.

\section*{Acknowledgment}


The research leading to these results has received funding from the European Community's Seventh Framework Programme (FP7/2007-2013) under grant agreement CRS-i number 318563. The research leading to this paper has been also supported in part by the Celtic-Plus project SHARING, project number C2012/1-8.



\end{document}